\documentclass[11pt,letterpaper]{article}

\usepackage[margin=1in]{geometry}
\usepackage[T1]{fontenc}
\usepackage{microtype}
\usepackage{parskip}
\usepackage{setspace}
\usepackage{multirow}
\usepackage[sc]{mathpazo}        % Palatino body + matching math
\usepackage[scale=0.95]{tgheros} % Helvetica/Arial-like for headings
\usepackage{sectsty}
\usepackage{graphicx}
\usepackage[table]{xcolor}
\usepackage{booktabs}
\usepackage{tabularx}
\usepackage{longtable}
\usepackage{array}
\usepackage[export]{adjustbox}   % <-- export enables valign keys with graphics
\usepackage{makecell}
\usepackage{float}

\usepackage{caption}
\usepackage[backend=biber,style=numeric,sorting=none,doi=true,url=true]{biblatex}
\usepackage{url}             % nicer URL line breaks
\usepackage[colorlinks=true]{hyperref}
\hypersetup{
  colorlinks=true,
  linkcolor=blue!50!black,
  urlcolor=blue!50!black,
  citecolor=blue!50!black,
  pdfauthor={Shams El-Adawy et al.},
  pdftitle={Categorization of Roles in the Quantum Industry}
}

\usepackage{titletoc}
\titlecontents{section}
  [0em]{\vspace{2pt}}{\contentslabel{1.5em}}{}{\titlerule*[.5pc]{.}\contentspage}
\titlecontents{subsection}
  [1.5em]{\vspace{1pt}}{\contentslabel{1.8em}}{}{\titlerule*[.5pc]{.}\contentspage}

\usepackage{enumitem}
\setlist{nosep,leftmargin=1.2em}

\usepackage[most]{tcolorbox}
\newtcolorbox{takeawaybox}{
  colback=blue!5!white, colframe=blue!40!black,
  boxrule=0.8pt, arc=2mm, left=6pt, right=6pt, top=6pt, bottom=6pt,
  title=\sffamily\bfseries Main Takeaways
}
\newtcolorbox{questionbox}{
  enhanced, breakable,
  colback=gray!8, colframe=gray!50,
  borderline west={3pt}{0pt}{gray!60},
  boxrule=0.5pt, arc=2mm,
  left=8pt, right=8pt, top=8pt, bottom=8pt,
  fonttitle=\sffamily\bfseries, coltitle=black,
  title={\sffamily\bfseries Key Question}
}

\AtBeginEnvironment{quote}{\small\itshape}

\usepackage{pagecolor}

\definecolor{brandnavy}{HTML}{243B7A}

\definecolor{giantsorange}{HTML}{F58F29}
\definecolor{lightorange}{HTML}{FCD1C5}
\definecolor{brick}{HTML}{AD2E24}
\definecolor{celestial}{HTML}{0576B3}
\definecolor{darkgreen}{HTML}{4E6E5D}

\sectionfont{\sffamily\bfseries\color{brandnavy}}
\subsectionfont{\sffamily\bfseries\large\color{brandnavy}}
\subsubsectionfont{\sffamily\bfseries\normalsize\color{brandnavy}}
\usepackage{titlesec}
\titleformat{\paragraph}[runin]{\sffamily\bfseries\color{brandnavy}}{\theparagraph}{0.5em}{}[]
\titleformat{\subparagraph}[runin]{\sffamily\bfseries\color{brandnavy}}{\thesubparagraph}{0.5em}{}[]

\arrayrulecolor{brandnavy!35} % soften vertical/horizontal rules a bit

\definecolor{tabheadbg}{RGB}{236, 240, 247} % very light navy tint
\definecolor{tabstripe}{RGB}{248, 250, 253} % ultra-light row stripe
\definecolor{tabskillbg}{RGB}{243, 246, 252} % background for the 4 skill columns

\newcolumntype{C}[1]{>{\centering\arraybackslash}p{#1}}
\newcolumntype{S}[1]{>{\columncolor{tabskillbg}\centering\arraybackslash}p{#1}}

\usepackage{fancyhdr}
\renewcommand{\headrulewidth}{0.2pt}

\makeatletter
\renewcommand{\headrule}{%
  \vspace{-3pt}\hbox to\headwidth{\color{brandnavy}\leaders\hrule height \headrulewidth\hfill}}
\makeatother

\newcolumntype{P}[1]{>{\raggedright\arraybackslash}p{#1}}

\providecommand{\seriesfontsize}{\normalsize}
\providecommand{\seriesitem}[3]{}
\providecommand{\seriesitemcurrent}[3]{}
\providecommand{\seriesstripe}[5]{}

\renewcommand{\seriesfontsize}{\large} % choose \normalsize, \large, or \Large
\newcommand{\serieslinesep}{12pt}       % vertical gap between list lines

\renewcommand{\seriesitem}[3]{%
  \noindent
  \begin{minipage}[t]{0.72\linewidth}
    {\sffamily\seriesfontsize\color{white} Report #1:\enspace #2}%
  \end{minipage}%
  \begin{minipage}[t]{0.28\linewidth}\raggedleft
    {\sffamily\seriesfontsize\color{white!85} #3}%
  \end{minipage}\par\vspace{\serieslinesep}%
}

\renewcommand{\seriesitemcurrent}[3]{\seriesitem{#1}{#2}{#3}}

\renewcommand{\seriesstripe}[5]{%
  \begin{tcolorbox}[enhanced, breakable,
    colback=brandnavy, colframe=brandnavy,
    boxrule=0pt, sharp corners,
    left=0pt, right=0pt, top=4pt, bottom=10pt,
    borderline south = {1pt}{0pt}{white!80}
  ]
    \noindent
    \begin{minipage}[t]{0.62\linewidth}
      {\sffamily\small\color{white!85}\MakeUppercase{#1}}%
    \end{minipage}%
    \begin{minipage}[t]{0.38\linewidth}\raggedleft
      {\sffamily\small\color{white!85} Publication date}%
    \end{minipage}

    \par\vspace{3pt}{\color{white!70}\rule{\linewidth}{0.4pt}}\vspace{4pt}

    \seriesitemcurrent{#2}{#4}{#3}

    {\sffamily\seriesfontsize\color{white!85} }\par\vspace{2pt}
    #5
  \end{tcolorbox}%
}

\newcommand{\upcomingseriesitems}{%
  \seriesitem{2}{Categorization of Roles in the Quantum Industry}{November 2025}%
  \seriesitem{3}{Profiles of Roles in the Quantum Industry}{TBD}%
}
\newcommand{\tocsection}[1]{%
  \phantomsection
  \addcontentsline{toc}{section}{#1}%
  \section*{#1}%
}

\newcommand{\logorowscale}{0.90}
\newcommand{\logospace}{1.2cm}

\title{%
  \vspace{-1.2cm}%
  \resizebox{\logorowscale\textwidth}{!}{%
    \makebox[\textwidth]{%
      \adjincludegraphics[valign=c,height=0.66cm]{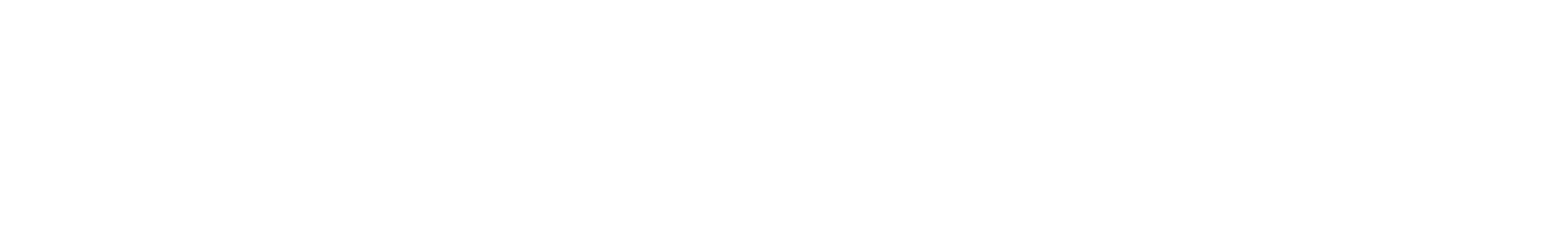}%
      \hspace{\logospace}%
      \adjincludegraphics[valign=c,height=0.60cm]{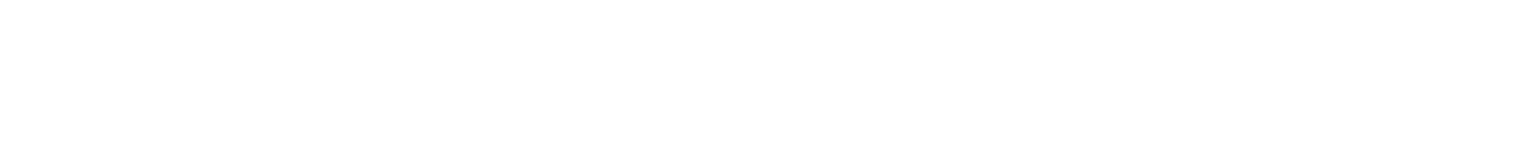}%
    }%
  }\\[1.2cm]
  {\sffamily\bfseries\Huge Categorization of Roles in the Quantum Industry}\\[0.6cm]
  {\sffamily\large Report 2 — November 2025}
}

\author{%
  \parbox{\textwidth}{\centering
    \begingroup
    \setlength{\tabcolsep}{18pt}%
    \begin{tabular*}{0.9\textwidth}{@{\extracolsep{\fill}} c c}
      \makecell{\textbf{Shams El-Adawy}\\ \textbf{Heather J. Lewandowski}\\[2pt]\small University of Colorado Boulder} &
      \makecell{\textbf{A.\,R. Pi\~{n}a}\\ \textbf{Benjamin M.\,Zwickl}\\[2pt]\small Rochester Institute of Technology}
    \end{tabular*}
    \endgroup
  }%
}

\date{}

\begin{document}

% Title page only (dark), then return to white 
\begingroup
  \pagecolor{brandnavy}\color{white}
  \hypersetup{linkcolor=cyan!25, urlcolor=cyan!25, citecolor=cyan!25}

  \maketitle

  % One-line, comment-free call avoids brace/mode errors
  \seriesstripe{Quantum Workforce Report Series}{1}{October 2025}{\href{https://doi.org/10.48550/arXiv.2510.12936}{Experimental Skills for Non-PhD Roles in the Quantum Industry}}{\upcomingseriesitems}

  \thispagestyle{empty}
\endgroup

%  Now normal white pages for the rest 
\clearpage
\nopagecolor
\color{black}
\hypersetup{linkcolor=blue!50!black, urlcolor=blue!50!black, citecolor=blue!50!black}

\tableofcontents
\thispagestyle{empty}
\clearpage

%  CONTENT 

\begin{figure}[h]
    \centering
    \includegraphics[width=.9\linewidth]{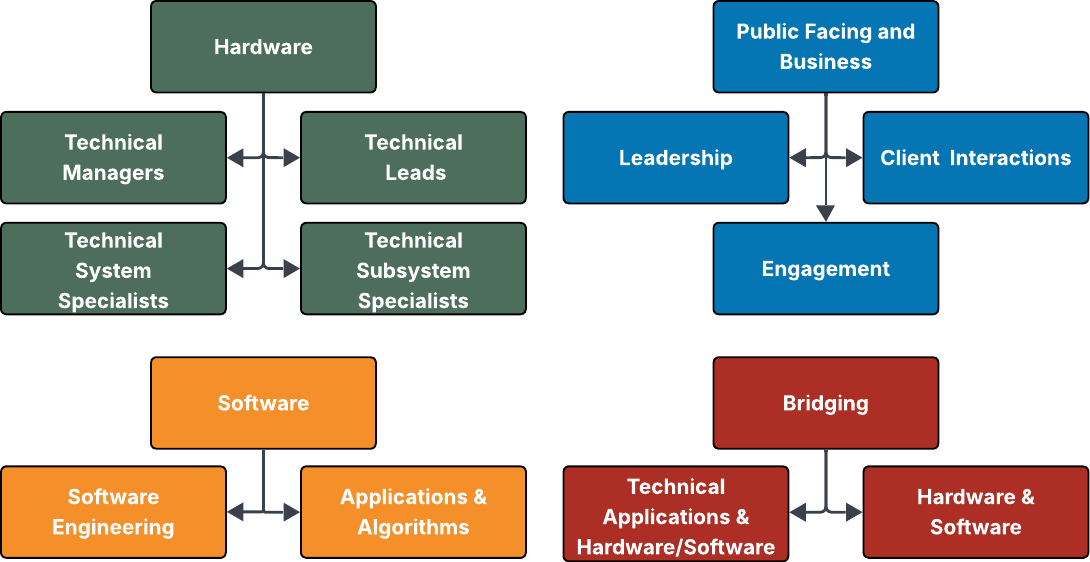}
    \caption{Summary of major categories and their subcategories for roles in the QISE industry.}
    \label{fig:summary_of_roles}
\end{figure}

\tocsection{Executive summary}
Continued growth of the quantum information science and engineering (QISE) industry has resulted in stakeholders spanning education \cite{el2025insights}, industry \cite{el2025industry}, and government \cite{CHIPS, NQI, NSP} seeking to better understand the workforce needs. 
This report presents a framework for the categorization of roles in the QISE industry based on 42 interviews of QISE professionals across 23 companies, as well as a description of the method used in the creation of this framework. 
The data included information on over 80 positions, which we have grouped into 29 roles spanning four primary categories: 
\begin{enumerate}
    \item \textbf{Hardware:} Roles focused on building, maintaining, and scaling quantum hardware systems.
    \item \textbf{Software:} Roles focused on designing, developing, and optimizing software for quantum systems and applications.
    \item \textbf{Bridging:} Roles focused on connecting teams within an organization (e.g., bridging the gap between technical applications and the underlying hardware or software)
    \item \textbf{Public Facing and Business:} Roles focused on business strategy, leadership, partnerships, public engagement, and government relations.
\end{enumerate}
For each primary category we provide an overview of what unites the roles within a category, a description of relevant subcategories (See Fig. \ref{fig:summary_of_roles}), and definitions of the individual roles (See Tab. \ref{summary_of_roles}). 
These roles serve as the basis upon which we generate profiles of these roles, which include information about role critical tasks, necessary knowledge and skills, and educational requirements. 
Our next report will present such profiles for each of the roles presented herein. 

\clearpage
\tocsection{Motivation}
Developing a better understanding of  the types and organization of roles within the QISE industry that is based on tasks, knowledge, skills, and abilities needed for those roles has significant potential implications for students, educators, and industry members. Knowing what different roles exist in the industry will allow instructors to better prepare students to fill those roles when they enter the workforce and  it will allow students to tailor their education to the types of careers they wish to pursue. 
Previous work has examined the knowledge, skills, and abilities needed for the workforce, but this work has some limitations.
The two primary data sources for more quantitative work in this area have been job ads \cite[e.g.,][]{goorney2025quantumtechnologyjobmarket, CQE_degrees_report} and survey data \cite[e.g.,][]{hughes2022, greinart2023_workforce}. 
Job postings are plentiful, and allow for data to be collected on a large number of positions at once, but are limited in their detail. 
Fixed option survey data is faster to collect, but can also be limited in detail; in the case of \textcite{hughes2022}, the multiple choice options presented for KSAs were derived from job postings. 
In both situations, the KSAs are usually short phrases and, significantly, do not connect the KSAs to  on-the-job tasks. 
There have also been qualitative data collection efforts in both US \cite{fox2020preparing} and EU \cite{Greinert2024} contexts. 
In particular, \textcite{Greinert2024} demonstrates how rich interview data can be with details regarding the knowledge and skills required for the quantum workforce in the EU context. 

Therefore, we address those limitations by creating a structure of roles based on interview data with employers in US based companies. 
More broadly, we aim to develop a shared language of job roles across disciplines, companies, areas of expertise, and sectors (e.g., industry, academia, and government) that will help generate a more cohesive QISE ecosystem and facilitate communication among members of the community. 
This shared language will also allow us to better communicate about future directions for QISE and track the workforce needs of the industry over time. 
A thorough categorization of roles in the QISE industry also provides a framework around which we can develop detailed profiles of occupations. 
These profiles, which will be the focus of our next report, will include a much more detailed breakdown of the tasks associated with different roles and the knowledge, skills, and abilities (KSAs) needed to perform those tasks. 

One significant challenge in developing a categorization framework for roles in the QISE industry is the degree to which position titles vary between companies in the industry. 
Although titles are a natural and productive starting point for categorizing positions, there are situations in which they can be misleading as a basis upon which to group positions. 
There are positions in the industry with very similar titles that are expected to perform a very different set of tasks on the job, requiring different KSAs. 
Similarly, there are positions with titles that are not similar at all, but are highly aligned on those other dimensions. 
These differences can lead to inconsistencies when trying to consider the function of a position, what they would do on the job (i.e., tasks), and what they would need to know (i.e., KSAs).

\begin{figure}
    \centering
    \includegraphics[width=0.9\linewidth]{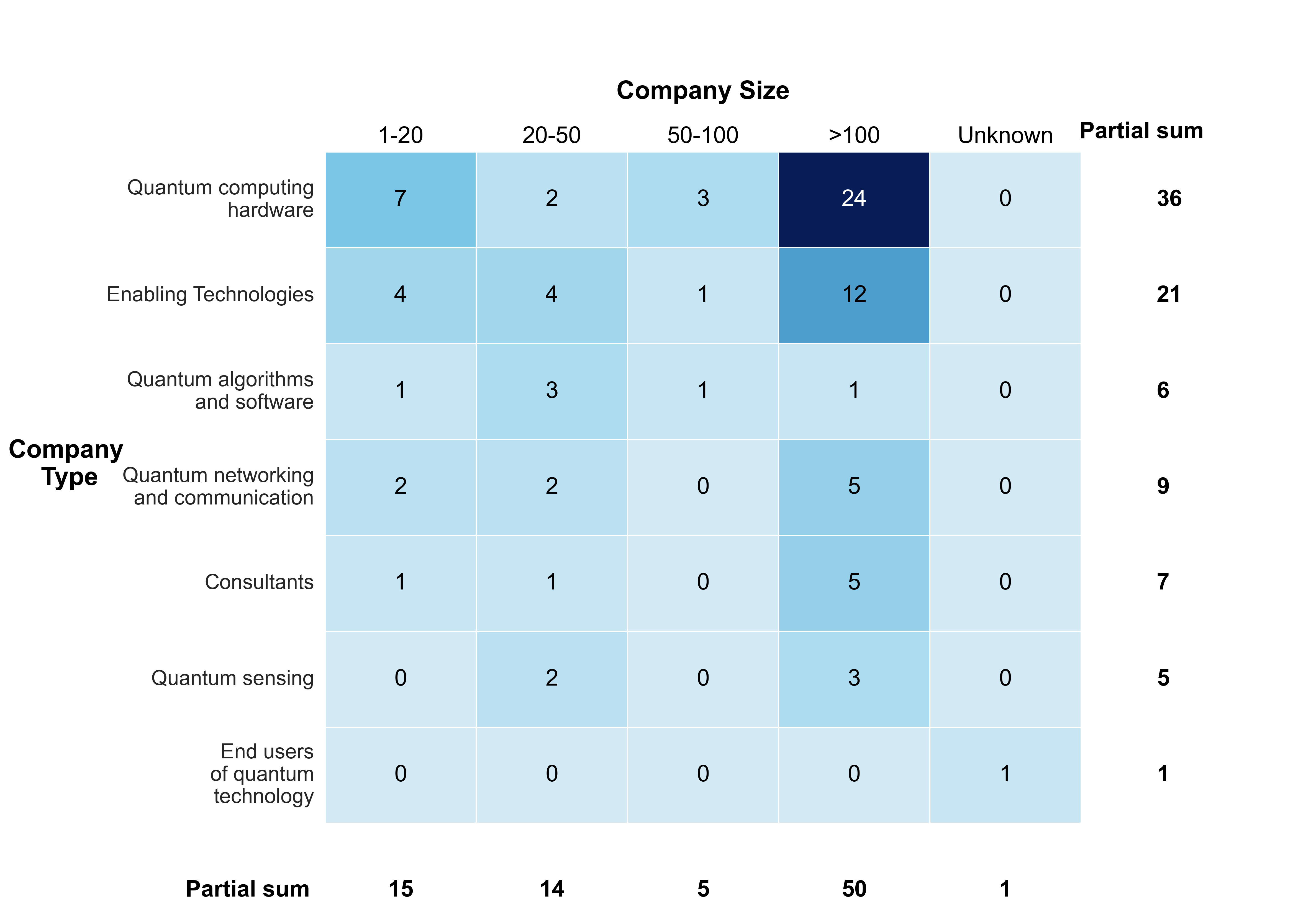}
    \caption{Distribution of interviewed companies by type and size.  We did 42 interviews with professionals at 23 distinct companies. We had a few interviewees from the same company and most companies self-reported participation in more than one type of activity. We adopted definitions consistent with prior literature \cite{fox2020preparing, el2025industry} for company types. Company size corresponds to the range of employees working in quantum-related technologies in these companies.}
    \label{fig:company_distribution}
\end{figure}

\tocsection{Data collection and analysis}
The data collected for this work were semi-structured interviews with professionals currently employed in the QISE industry. 
These individuals worked at companies spanning most sectors of the industry, however some are more represented than others (see Figure \ref{fig:company_distribution}). 
The interviews were conducted with two protocols, both of which included questions about the tasks, knowledge, skills, abilities, and educational requirements associated with one or more quantum-related positions at the participants' company. 
One protocol was intended for managers who could speak to a variety of positions within their company and the other was intended for employees  and focused on their specific position. 
Although the interviews did include other topics related to the QISE industry, these were the most pertinent questions in categorizing positions into the roles discussed here. 

\begin{questionbox}
\textbf{What is a structure of roles within the QISE industry based on tasks and KSAs that can be used to develop usable profiles for students, educators, and industry professionals?}
\end{questionbox}

The analysis we did to address this question happened in two primary stages. 
We started by considering position titles and education requirements. 
As discussed in the previous section, however, titles proved to be a sometimes misleading criterion for categorization. 
For example, take two positions, which based on titles alone, one might think are similar: \textit{Quantum Solutions Scientist} and \textit{Quantum Applications Scientist}. 
The former is a software role focused on design and develop of code packages directly related to quantum computing.
The latter facilitates the implementation of their company’s products in the customer’s context (e.g., enabling technologies). They combine technical knowledge with sales and relationship building skills. 

Two other positions such as \textit{Quantum Device Measurement Scientist} and \textit{Lab Engineer} have more distinct titles. 
However, both of these are positions perform device characterization by conducting specialized physical measurements (e.g., optical or  RF) of quantum systems and perform analysis of the generated data, which they use to make determinations about a device.

Education posed similar limitations as a basis for categorization. 
The field is highly interdisciplinary and positions can often be filled by individuals with education from a variety of fields. 
There were also positions for which any range of education from bachelor's through PhD was appropriate. 
Education requirements alone also do not account for the role of prior work experience in position requirements (e.g., a position requiring a bachelor's and four years relevant experience). 

Due to these emergent limitations, we adjusted the groups based on the overlap of tasks making considerations for the KSAs in any cases where tasks alone resulted in some ambiguity. 
In many cases, the consideration of the additional information led to some re-sorting. 
For example, \textit{Quantum Systems Engineer} and \textit{Quantum Applications Scientist} were grouped because both positions focus on integrating company products into client systems. 
Another example is the \textit{AMO Theorist} and \textit{Photonics Engineer}, which were grouped together due to a shared focus on computational modeling of quantum systems. 

In a few cases however, this additional information confirmed the initial title-based categorization. 
Positions such as \textit{Product Manager} and \textit{Project Manager}  proved to have tasks and KSAs in common. 
Another example of this confirmation was the alignment of tasks for positions like \textit{Software Developer}, \textit{Software Engineer}, and \textit{Scientific Software Engineer}. 

We followed up this categorization of positions into roles by grouping the roles by their primary function within an organization. 
\begin{itemize}
    \item Hardware roles engage in hands-on work designing, developing, refining, and manufacturing hardware for quantum technologies.
    \item Software roles engage in the work of designing, developing, and optimizing software for quantum systems and applications. 
    \item Bridging roles focus on connecting teams within an organization (e.g., bridging the gap between technical applications and the underlying hardware or software). 
    \item Public facing and business roles are connected by their shared responsibilities regarding engaging with the public or clients.
\end{itemize}

\newpage
\tocsection{Results}
Categorization of 81 positions recoded across 42 interviews resulted in 29 roles. 
A summary of the roles found in each of the four primary categories (hardware, software, bridging, and public facing \& business) can be found in Table \ref{summary_of_roles}.
A detailed breakdown of the sub-categories within each of the primary categories, as well as definitions of each role, will be provided in the following subsections. 

\begin{table}[h]
\caption{The four primary categories into which roles were sorted, as well as the codes and titles for each individual role.}
\label{summary_of_roles}
\begin{adjustbox}{max width=\textwidth, center}
\centering
\begin{tabularx}{.97225\textwidth}{|cll|}
\hline
\rowcolor{tabheadbg}
\multicolumn{1}{|c}{\textbf{\textcolor{brandnavy}{Category}}}                & \textbf{\textcolor{brandnavy}{Code}} & \textbf{\textcolor{brandnavy}{Role}}                                                \\\hline
\multirow{15}{*}{\textbf{\textcolor{darkgreen}{Hardware}}}               & H1.1 & Senior Scientists                                           \\
                                            & H1.2 & Engineering Managers                                        \\
                                            & H2.1 & Commercialization Leads                                     \\
                                            & H2.2 & Systems Engineers                                            \\
                                            & H3.1 & Experimental Scientists                                     \\
                                            & H3.2 & Quantum Hardware System Engineers                           \\
                                            & H3.3 & Field Deployment Engineers                                 \\
                                            & H4.1 & Superconducting Quantum Engineers                           \\
                                            & H4.2 & Device Characterization \& Measurement Specialists \\
                                            & H4.3 & Design Engineer, EE Circuits, RF Specialist                 \\
                                            & H4.4 & Optics \& Photonics Experiment Specialists                 \\
                                            & H4.5 & Optics \& Photonics Assembly Specialists                   \\
                                            & H4.6 & Cryogenics Specialists                                      \\
                                            & H4.7 & Nano/Microscale Specialists                                       \\
                                            & H4.8 & Lab and Construction Technicians                            \\\hline
\multirow{4}{*}{\textbf{\textcolor{giantsorange}{Software}}}                   & S1.1 & Traditional Software Engineers                              \\
                                            & S1.2 & Quantum Software Engineers                                  \\
                                            & S2.1 & Quantum Information Science Algorithms Theorists            \\
                                            & S2.2 & Quantum Algorithms Programmers                              \\\hline
\multirow{4}{*}{\textbf{\textcolor{brick}{Bridging}}}                   & B1.1 & Quantum Software Application Developers \& Trainers        \\
                                            & B1.2 & Quantum Technology End Users                                \\
                                            & B2.1 & Device \& System Hardware Computational Scientists         \\
                                            & B2.2 & Quantum Computer Operators                                  \\ \hline
\multirow{6}{*}{\textbf{\textcolor{celestial}{Public Facing \& Business}}} & P1.1 & Company Executives                                          \\
                                            & P1.2 & Project Overseers                                           \\
                                            & P2.1 & Hardware Applications \& Technical Sales Specialists       \\
                                            & P2.2 & Business \& Partnerships Specialists                       \\
                                            & P3.1 & Education Advocates                                          \\
                                            & P3.2 & Government-Industry Advocate \\ \hline                              
\end{tabularx}
\end{adjustbox}
\end{table}

\clearpage
\subsection*{\textcolor{darkgreen}{H: Hardware Roles}}
\addcontentsline{toc}{subsection}{H: Hardware Roles}
The hardware category consists of roles focused on building, maintaining, and scaling quantum hardware systems. Within the hardware category, roles were grouped by leadership responsibilities and the breadth of technical expertise (e.g., does the role require specialization for a specific subsystem or expertise spanning several subsystems and their integration).

\begin{figure}[h]
    \centering
    \includegraphics[width=\linewidth]{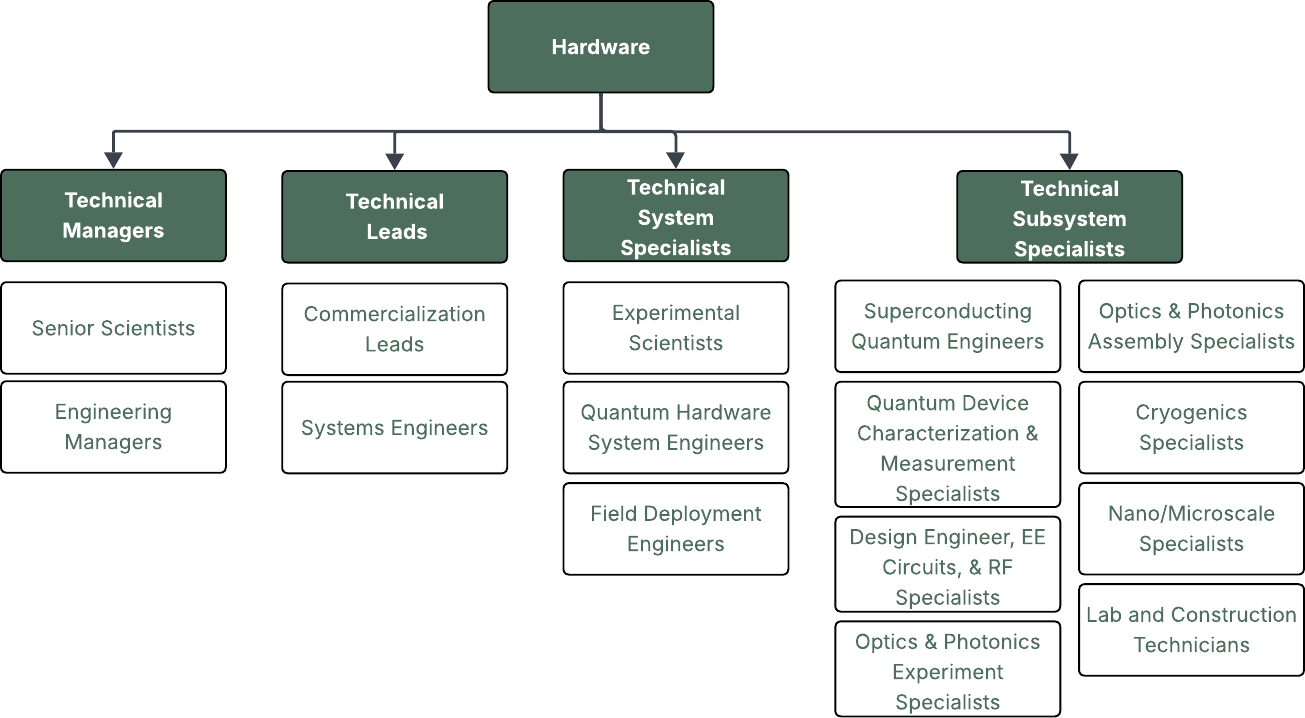}
    \caption{The structure of the roles in the hardware category.}
    \label{fig:placeholder}
\end{figure}

\subsubsection*{{\textcolor{darkgreen}{H1: Technical Managers}}}
\addcontentsline{toc}{subsubsection}{H1: Technical Managers}
Technical managers fill a dual role, bringing significant technical expertise and the ability to manage and coordinate teams for a concentrated hardware development effort. Roles focused on managing people and coordinating hardware development efforts. The breath of responsibilities requires broad technical knowledge. 

\begin{itemize}
    \item \textbf{H1.1 Senior Scientists:} Work in areas closely connected to fundamental research, they oversee technical projects and manage people.
    \item \textbf{H1.2 Engineering Managers:} Oversee teams of engineers building hardware systems.
\end{itemize}

\subsubsection*{{\textcolor{darkgreen}{H2: Technical Leads}}}
\addcontentsline{toc}{subsubsection}{H2: Technical Leads}
Similar to the technical managers, the technical leads bring significant technical expertise and the ability to manage and lead different hardware development efforts, but at a smaller scale than the technical manager. Where the technical managers coordinate teams, the technical leads would be the heads of one of those teams, focused on leading specific experiments or product development efforts. 
 
\begin{itemize}
    \item \textbf{H2.1 Commercialization Leads:} Lead a team and drive product development, commercialization, and applications. They coordinate and contribute to the design, construction, and operation of quantum technologies. 

    \item \textbf{H2.2 Systems Engineers:} Perform traditional systems engineering in a quantum setting. For example, when building a quantum computer, they must understand all of the systems and sub-systems, how they interact with one another, how they interface with relevant software, and how they are impacted by environmental factors.

\end{itemize}

\subsubsection*{{\textcolor{darkgreen}{H3: Technical System Specialists}}}
\addcontentsline{toc}{subsubsection}{H3: Technical System Specialists}
Hold domain expertise of hardware systems and are able to work with a variety of subsystems necessary to develop different quantum technologies.  

\begin{itemize}
    \item \textbf{H3.1 Experimental Scientists:} Plan and run experiments (e.g., doing quantum optics experiments or device testing), perform analysis, and report results either internally (e.g., report) or externally (e.g., publications).
    \item \textbf{H3.2 Quantum Hardware System Engineers:} Turn research results into devices. They design hardware subsystems operating within practical constraints and integrate them into larger systems.
    \item \textbf{H3.3 Field Deployment Engineers:} Deploy quantum technology products offsite and ensure the hardware operates as expected in the customer’s environment. Their main focus is to ensure the product is functioning correctly during offsite deployment.
\end{itemize}

\subsubsection*{{\textcolor{darkgreen}{H4: Technical Subsystem Specialists}}}
\addcontentsline{toc}{subsubsection}{H4: Technical Subsystem Specialists}
Technical subsystem specialists are roles that have a specific expertise or skill-set related to, or enacted in support of, a subsystem of quantum technology development or manufacturing effort.

\begin{itemize}
    \item \textbf{H4.1 Superconducting Quantum Engineers:} Model, design, fabricate, operate, and analyze superconducting qubits and circuits.
    \item \textbf{H4.2 Quantum Device characterization and measurement specialists:} Perform device characterization by conducting specialized physical measurements (e.g., optical or  RF) of quantum systems and by performing analysis of the generated data. They use these to make determinations about device performance. 
    \item \textbf{H4.3 Design Engineer, EE Circuits, RF Specialists:} Develop and test classical electronic circuits for applications in quantum technologies. 
    \item \textbf{H4.4 Optics and Photonics Experiment Specialists:} Support experiments, such as maintaining optical systems and performing some data collection and analysis. 
    \item \textbf{H4.5 Optics and photonics assembly specialists:} Perform assembly, testing, and quality control for optical/photonic systems
    \item \textbf{H4.6 Cryogenics Specialist:} Develop systems to test quantum devices at cryogenic temperatures.
    \item \textbf{H4.7 Nano/Microscale Specialists:} Fabricate nano/microscale devices, typically in a cleanroom setting. They conduct nano/microscale patterning, utilizing various techniques dependent on the architecture of the devices. They perform characterization of fabricated nano/microscale devices.
    \item \textbf{H4.8 Lab and Construction Technicians:} Build and maintain facilities for the fabrication and testing of quantum devices. This includes planning and overseeing the construction of cleanroom facilities, maintaining mechanical connections, and general electrical or plumbing work. 
\end{itemize}

\clearpage
\subsection*{\textcolor{giantsorange}{S: Software Roles}}
\addcontentsline{toc}{subsection}{S: Software Roles}
The software category consists of roles focused on designing, developing, and optimizing software for quantum systems and applications.  Within the software category, there are roles more focused on developing new software utilizing primarily classical methods and roles focused on developing algorithms for, or running algorithms on, quantum hardware. 
\begin{figure}[h]
    \centering
    \includegraphics[width=0.35\linewidth]{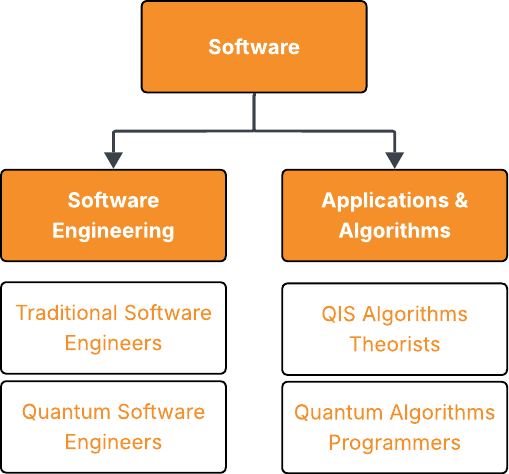}
    \caption{The structure of roles in the software category.}
    \label{fig:placeholder}
\end{figure}

\subsubsection*{\textcolor{giantsorange}{S1: Software Engineering}}
Software engineering roles research, design, and develop computer and network software or specialized utility programs. The dividing line between the following two roles are the focus and utility of those programs. 
\addcontentsline{toc}{subsubsection}{S1: Software Engineering}
\begin{itemize}
    \item \textbf{S1.1 Traditional Software Engineers:} Build classical software systems that support or integrate with quantum workflows.
    \item \textbf{S1.2 Quantum Software Engineers:} Design and develop software directly related to quantum computing.
\end{itemize}

\subsubsection*{\textcolor{giantsorange}{S2: Applications and Algorithms}}
Applications and algorithms specialists develop, implement, and optimize algorithms for quantum applications. 
\addcontentsline{toc}{subsubsection}{S2: Applications and Algorithms}
\begin{itemize}
    \item \textbf{S2.1 Quantum Information Science Algorithms Theorists:} Utilize theory for conceptual design, mathematical formulation, and optimization of quantum algorithms.
    \item \textbf{S2.2 Quantum Algorithms Programmers:} Focus on implementing, testing, and optimizing quantum algorithms on quantum computing platforms.
\end{itemize}

\clearpage
\subsection*{\textcolor{brick}{B: Bridging Roles}}
\addcontentsline{toc}{subsection}{B: Bridging Roles}
The bridging category consists of roles that work to facilitate communication and collaboration between different roles within a company. Bridging roles require expertise in more than one domain for the purpose of ``bridging the gap'' between those domains to achieve a goal. 
\begin{figure}[h]
    \centering
    \includegraphics[width=0.35\linewidth]{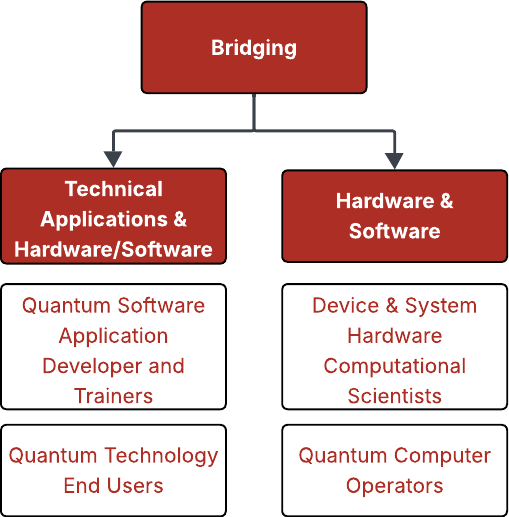}
    \caption{The structure of roles in the bridging category.}
    \label{fig:placeholder}
\end{figure}
\subsubsection*{\textcolor{brick}{B1: Bridging Technical Applications and Software/Hardware}}
Bridging technical applications with quantum hardware/software involves understanding the applications of quantum technologies and finding ways to implement specific quantum hardware or software in pursuit of that application. 
\addcontentsline{toc}{subsubsection}{B1: Bridging Technical Applications and Software/Hardware}
\begin{itemize}
    \item \textbf{B1.1 Quantum Software Application Developers and Trainers:} Develop real-world applications that make use of quantum software and apply quantum algorithms. They also play a key role in training customers to effectively use quantum software tools.
    \item \textbf{B1.2 Quantum Technology End Users:} Explore ways to leverage quantum technologies to domain-specific problems for their company, whose primary business is outside of QISE. 
\end{itemize}

\subsubsection*{\textcolor{brick}{B2: Bridging Hardware and Software}}
\addcontentsline{toc}{subsubsection}{B2: Bridging Hardware and Software}
Roles bridging hardware and software use software or computational tools to model quantum systems and make actionable recommendations to hardware specialists or work on the code that controls and runs on quantum hardware.  In either case, these roles require a mix of hardware and software domain expertise. 
\begin{itemize}
    \item \textbf{B2.1 Device and System Hardware Computational Scientists:} Computationally model quantum devices and systems to support design, validation, and optimization.
    \item \textbf{B2.2 Quantum Computer Operators:} Operate, maintain, and schedule software tasks to run on quantum hardware. 

\end{itemize}

\clearpage
\subsection*{\textcolor{celestial}{P: Public Facing and Business Roles}}
\addcontentsline{toc}{subsection}{P: Public Facing and Business Roles}
The public-facing and business category consists of roles focused on business strategy, leadership, partnerships, public engagement, and government relations.

\begin{figure}[h]
    \centering
    \includegraphics[width=0.5\linewidth]{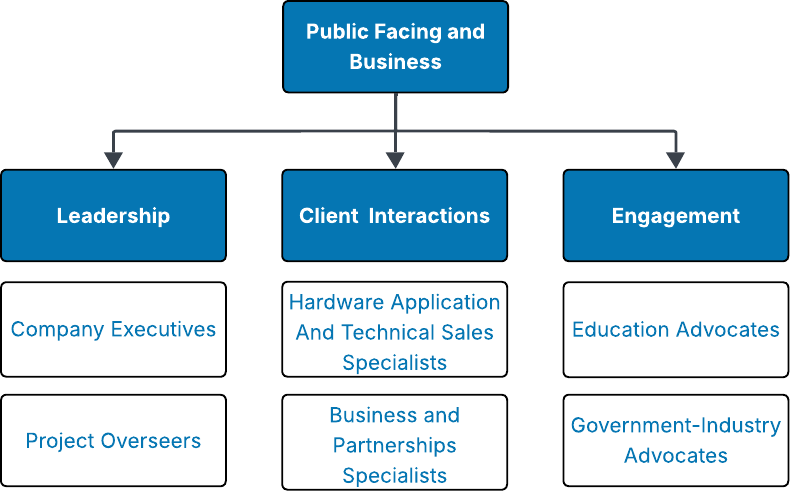}
    \caption{The structure of roles in the public facing and business category.}
    \label{fig:placeholder}
\end{figure}
\subsubsection*{\textcolor{celestial}{P1: Leadership}}
\addcontentsline{toc}{subsubsection}{P1: Leadership}
Leadership roles are focused on determining company direction and management of the various projects and initiatives the company undertakes. These leadership roles are typically focused on big picture oversight and may not require as much technical expertise as the managers or leads within the Hardware and Software categories. 
\begin{itemize}
    \item \textbf{P1.1 Company Executives:} Make strategic decisions and are responsible for vision, growth, and overall management of the organization.
    \item \textbf{P1.2 Project Overseers:} Manage timelines, resources, and deliverables for projects involving quantum technologies.
\end{itemize}

\subsubsection*{\textcolor{celestial}{P2: Client Interactions}}
\addcontentsline{toc}{subsubsection}{P2: Client Interactions}
Client interactions includes roles focused on coordinating with groups outside of the company for sales or the implementation of products in customer contexts. 
\begin{itemize}
    \item \textbf{P2.1 Hardware Application \& Technical Sales Specialists:} Understand customer needs and translate them into solutions using quantum hardware. They facilitate the implementation of their company’s products in the customer’s context (e.g., enabling technologies). They combine technical knowledge with sales and relationship building skills. 
    \item \textbf{P2.2 Business and Partnerships Specialists:} Focus on strategic business development, partnerships, and market expansion in the quantum space.
\end{itemize}

\subsubsection*{\textcolor{celestial}{P3: Engagement}}
\addcontentsline{toc}{subsubsection}{P3: Engagement}
The engagement category consists of roles focused on interfacing with the public or government entities for the purposes of education and advocacy for and about the QISE industry. 
\begin{itemize}
    \item \textbf{P3.1 Education Advocates:} Focus on outreach and education to increase awareness and understanding of quantum technologies.
    \item \textbf{P3.2 Government-Industry Advocates:} Advocate for industry needs with government agencies. They synthesize technical expertise provided by scientists in their company to write proposals and manage projects across the company. 

\end{itemize}

\subsection*{Limitations}

Our findings should be interpreted with the following limitations in mind:

\begin{itemize}
  \item \textit{Sample and scope:} We interviewed 42 professionals at 23 companies. Thus, the sample is not intended to be representative of all roles in the quantum industry.
  \item \textit{Level of detail of responses:} Across the different interviews, there was  variation in the level of detail participants discussed for different positions. As a result, some of the definitions presented here may be more refined than others. 
  \item \textit{Time-bounded snapshot:} The report presents the first set of findings based on interviews conducted with quantum industry professionals in 2025. Workforce needs may evolve as quantum technologies becomes more mature.
\end{itemize}

\subsection*{Future work}

Building on these initial results and addressing the limitations above, we plan to:
\begin{itemize}
   \item Develop profiles of each of these roles incorporating information on tasks, KSAs, and educational requirements.
  \item Conduct additional interviews and target roles for which we have less data.
\end{itemize}
\newpage
\tocsection{Acknowledgments}
% \addcontentsline{toc}{section}{Acknowledgments}
Thank you to the quantum industry managers and employees who participated in our interviews. 

This material is based on work supported by the National
Science Foundation under Grant Nos. PHY-2333073 and PHY-2333074.

This material is also based on work supported by the Army Research Office and was accomplished under Award Number: W911NF-24-1-0132. The views and conclusions contained in this document are those of the authors and should not be interpreted as representing the official policies, either expressed or implied, of the Army Research Office or the U.S. Government. The U.S. Government is authorized to reproduce and distribute reprints for Government purposes notwithstanding any copyright notation herein.

\vspace{0.9\baselineskip}
{\sffamily\bfseries\color{brandnavy} About the project}\\
This report is part of a collaborative project between researchers at the University of Colorado Boulder \& Rochester Institute of Technology to advance quantum information science education and strengthen the quantum workforce.  
Learn more about the broader effort at
\href{https://www.rit.edu/quantumeducationandworkforce/}{rit.edu/quantumeducationandworkforce}.

\vspace{0.9\baselineskip}
{\sffamily\bfseries\color{brandnavy} Contact information}\\
For questions about this report or the broader project, please contact the principal investigators:

\noindent\hfill
\begin{minipage}{0.9\linewidth}
\centering
\begin{tabular}{@{}p{0.45\linewidth}p{0.45\linewidth}@{}}
Heather J.~Lewandowski & Benjamin M.~Zwickl \\
University of Colorado Boulder & Rochester Institute of Technology \\
\href{mailto:lewandoh@colorado.edu}{lewandoh@colorado.edu} &
\href{mailto:bmzsps@rit.edu}{bmzsps@rit.edu} \\
\end{tabular}
\end{minipage}
\hfill\mbox{}

\vspace{0.9\baselineskip}
{\sffamily\bfseries\color{brandnavy} Suggested citation}\\
A.\,R. Pi\~na, Shams El-Adawy, H. J. Lewandowski, \& Benjamin M.\,Zwickl (November 2025).
\emph{Categorization of Roles in the Quantum Industry}
(Quantum Workforce Report Series, Report 2). University of Colorado Boulder \& Rochester Institute of Technology.

\newpage

% \bibliographystyle{unsrt}
% \bibliography{references}
\printbibliography

@article{el2025insights,
  title = {Insights from educators on building a more cohesive quantum information science and engineering education ecosystem},
  author = {El-Adawy, Shams and Pi\~na, A. R. and Zwickl, Benjamin M. and Lewandowski, H. J.},
  journal = {Phys. Rev. Phys. Educ. Res.},
  volume = {21},
  issue = {2},
  pages = {020144},
  numpages = {20},
  year = {2025},
  month = {Nov},
  publisher = {American Physical Society},
  doi = {10.1103/tfmb-hnvz},
  url = {https://link.aps.org/doi/10.1103/tfmb-hnvz}
}

@inproceedings{el2025industry, Author = "Shams El-Adawy and A. R. Piña and Benjamin Zwickl and Heather J. Lewandowski", Title = {Industry Perspectives on Projected Quantum Workforce Needs}, BookTitle = {Physics Education Research Conference 2025}, Pages = {148-153}, Address = {Washington, DC}, Series = {PER Conference}, Month = {August 6-7}, Year = {2025} }

@article{fox2020preparing,
  title     = {Preparing for the quantum revolution: What is the role of higher education?},
  author    = {Fox, Michael FJ and Zwickl, Benjamin M and Lewandowski, Heather J},
  journal   = {Physical Review Physics Education Research},
  volume    = {16},
  number    = {2},
  pages     = {020131},
  year      = {2020},
  publisher = {APS},
  url       = {https://doi.org/10.1103/PhysRevPhysEducRes.16.020131}
}

@misc{CQE_degrees_report, 
    title={Quantum Tech Job Accessibility: 2024 CQE Study}, 
    url={https://chicagoquantum.org/degreereports}, 
    journal={Chicago Quantum Exchange}, 
    author={Fore, Meredith and Gillespie, Becky Beaupre and Prich, Tyler}, 
    year={2024}}

@misc{goorney2025quantumtechnologyjobmarket,
      title={The Quantum Technology Job Market: Data Driven Analysis of 3641 Job Posts}, 
      author={Simon Goorney and Eleni Karydi and Borja Muñoz and Otto Santesson and Zeki Can Seskir and Ana Alina Tudoran and Jacob Sherson},
      year={2025},
      eprint={2503.19004},
      archivePrefix={arXiv},
      primaryClass={physics.ed-ph},
      url={https://arxiv.org/abs/2503.19004}, 
}

@ARTICLE{hughes2022,
  author={Hughes, Ciaran and Finke, Doug and German, Dan-Adrian and Merzbacher, Celia and Vora, Patrick M. and Lewandowski, H. J.},
  journal={IEEE Transactions on Education}, 
  title={Assessing the Needs of the Quantum Industry}, 
  year={2022},
  volume={65},
  number={4},
  pages={592-601},
  doi={10.1109/TE.2022.3153841}}

@article{greinart2023_workforce,
  title = {Future quantum workforce: Competences, requirements, and forecasts},
  author = {Greinert, Franziska and M\"uller, Rainer and Bitzenbauer, Philipp and Ubben, Malte S. and Weber, Kim-Alessandro},
  journal = {Phys. Rev. Phys. Educ. Res.},
  volume = {19},
  issue = {1},
  pages = {010137},
  numpages = {19},
  year = {2023},
  month = {Jun},
  publisher = {American Physical Society},
  doi = {10.1103/PhysRevPhysEducRes.19.010137},
  url = {https://link.aps.org/doi/10.1103/PhysRevPhysEducRes.19.010137}
}

@article{Greinert2024,
   author = {Franziska Greinert and Malte S Ubben and Ismet N Dogan and Dagmar Hilfert-Rüppell and Rainer Müller},
   doi = {10.1140/epjqt/s40507-024-00294-2},
   issn = {2196-0763},
   issue = {1},
   journal = {EPJ Quantum Technology},
   pages = {82},
   title = {Advancing quantum technology workforce: industry insights into qualification and training needs},
   volume = {11},
   url = {https://doi.org/10.1140/epjqt/s40507-024-00294-2},
   year = {2024}
}

@misc{CHIPS,
	type = {legislation},
	title = {H.{R}.4346 - 117th {Congress} (2021-2022): {Chips} and {Science} {Act}},
	copyright = {Text is government work},
	shorttitle = {H.{R}.4346 - 117th {Congress} (2021-2022)},
	url = {http://www.congress.gov/},
	urldate = {2023-06-01},
	author = {Rep. Ryan, Tim [D-OH-13]},
	month = aug,
	year = {2022},
	note = {Archive Location: 08/09/2022},
}

@misc{NSP,
	title = {Quantum Information Science and Technology Workforce Development National Strategic Plan},
	url = {https://www.quantum.gov/wp-content/uploads/2022/02/QIST-Natl-Workforce-Plan.pdf},
	pages = {34},
	institution = {National Science and Technology Council},
	author = {Committee on Science of the National Science \{and\} Technology Council},
	year = {2022},
}

@misc{NQI,
	type = {legislation},
	title = {H.{R}.6227 - 115th {Congress} (2017-2018): {National} {Quantum} {Initiative} {Act}},
	copyright = {Text is government work},
	shorttitle = {H.{R}.6227 - 115th {Congress} (2017-2018)},
	url = {https://www.congress.gov/bill/115th-congress/house-bill/6227},
	urldate = {2023-01-17},
	author = {Smith, Lamar},
	month = dec,
	year = {2018},
	note = {Archive Location: 2017/2018},
}
\end{document}